\documentclass[aps,prl,twocolumn,preprintnumbers,showpacs,nofootinbib]{revtex4}

\usepackage{graphicx}% Include figure files
\usepackage{dcolumn}% Align table columns on decimal point
\usepackage{bm}% bold math
\begin{document}
\newcommand{\Od}{{\cal O}}
\newcommand{\lsim}   {\mathrel{\mathop{\kern 0pt \rlap
  {\raise.2ex\hbox{$<$}}}
  \lower.9ex\hbox{\kern-.190em $\sim$}}}
\newcommand{\gsim}   {\mathrel{\mathop{\kern 0pt \rlap
  {\raise.2ex\hbox{$>$}}}
  \lower.9ex\hbox{\kern-.190em $\sim$}}}

%\preprint{APS/123-QED}

\title{Brane-world dark matter}% Force line breaks with \\

\author{J.A.R. Cembranos}
%\email{cembra@fis.ucm.es}
\author{A. Dobado}%
%\email{malcon@fis.ucm.es}
\author{A.L. Maroto}
%\email{maroto@fis.ucm.es}
\affiliation{Departamento de  F\'{\i}sica Te\'orica,
 Universidad Complutense de
  Madrid, 28040 Madrid, Spain}%

\date{\today}% It is always \today, today,

\begin{abstract}
We show that, in the context of brane-world scenarios with low
tension $\tau=f^4$, massive brane fluctuations are natural
dark matter candidates. We calculate the present
 abundances for both hot(warm) and cold
branons in terms of the branon mass $M$ and
the tension scale $f$. The results are compared
with the current experimental bounds on these parameters.
 We also study the prospects for their detection
in direct search experiments and comment
on their characteristic signals in the indirect ones.
 \\
\end{abstract}

\maketitle

One of the most important open problems in astrophysics and
cosmology is to identify the nature of dark matter. It has
been known for a long time that the luminous matter observed
in spiral galaxies is insufficient to explain their rotation
curves. The existence of dark halos 
was  proposed as a possible solution for the discrepancy 
(see \cite{Binney} and references therein),
although at present, 
numerical simulations of the formation of such halos appear
inconsistent with observations. On the other hand, different
estimations of the total matter density in the universe
from large scale motions, virial masses
or cluster abundances, and the more recent Type Ia supernovae  
and CMB anisotropies observations agree in a value
$\Omega_M=0.27\pm 0.04$ \cite{PDG,WMAP}, which is much larger 
than the
value of the total luminous mass density in the universe
$\Omega_{lum}h=0.006 - 0.002$. In addition,
the big bang nucleosynthesis (BBN) and WMAP results for the 
total baryonic content
 $\Omega_B h^2=0.0224\pm 0.0009$
imply that most of the matter in the universe is
dark and non-baryonic (see \cite{PDG,WMAP,Kamion} and 
references therein).

A possible explanation of this puzzle is that the dominant
component of dark matter consists of some non-relativistic (cold)
stable and weakly interacting massive particles (WIMP) which
decoupled from radiation early enough so that their relic
abundances are important today. The possibility that the universe is
dominated by hot dark matter seems to conflict with numerical
simulations of structure formation. Thus, the only potential
candidates within the known particles would be  massive neutrinos.
However a detailed analysis has excluded the three light 
and even one additional heavy fourth generation 
of Majorana or Dirac neutrinos.
This fact has led to the search of cold dark matter candidates
beyond the Standard Model (SM)  \cite{Kamion}.

There are two main such particles studied in the literature. On
one hand axions which strictly speaking cannot be considered as
 WIMP's since they are very light and produced non-thermally. On the
other hand we have the lightest supersymmetric particle, which can
be identified with a neutralino in most of the supersymmetric
models. The latter is probably the most studied and best
theoretically motivated dark matter candidate \cite{Kamion}.
However the large number of free parameters in supersymmetric
theories make their predictions extremely model dependent. More
recently, the existence of large extra dimensions has been proposed
as a new setting for a possible solution to the hierarchy problem
\cite{ADD}.  In this scenario, the SM fields are forced to
live on a three-dimensional hypersurface (brane) whereas gravity
is able to propagate on the higher $D=4+N$ dimensional bulk space.
 In this  Brane World scenario
(BWS) the fundamental scale of gravity is not the Planck scale
$M_P$ but another scale $M_D$. Although in the original  
ADD model \cite{ADD}, $M_D$ is supposed to be not too much
larger than the electroweak scale, recently brane 
 cosmology models  
have arised in which $M_D$ is  much larger than the 
TeV  (\cite{Langlois} and references therein). In this work we will
 consider the case of a general BWS without assuming any particular
 value for $M_D$. 

In these models gravitons 
propagating through the bulk space give rise to a
Kaluza-Klein (KK) tower of massive gravitons  on the
brane. These KK gravitons couple to the energy-momentum tensor of
the SM fields $T_{SM}^{\mu\nu}$ and could be produced under the
appropriate circumstances as real or virtual particles.
Another important effect that is expected in the BWS is the
presence of brane fluctuations since rigid objects do not exist in
relativistic theories. In other words the brane should have some
finite tension $\tau=f^4$. When these oscillations are taken into
account two new effects appear \cite{GB}. First of all, we have to
introduce new fields, which for a homogeneous extra space, essentially
represent the position of the brane in the bulk space $(x^{\mu},
y^\alpha \simeq \pi^\alpha(x)/f^2)$. The $\pi^\alpha(x)$ 
fields are the
Goldstone bosons (GB) corresponding to the spontaneous symmetry
breaking (SSB) of the translational invariance produced by the
presence of the brane (branons). It has been shown \cite{GB} 
that when these
branons are properly taken into account, the coupling of the SM
particles to any bulk field is exponentially suppressed by a
factor $\exp(-M^2_{KK}M_D^2/(8 \pi^2f^4))$, where $M_{KK}$ is the
mass of the corresponding KK mode. As a consequence, if the tension
scale $f$ is much smaller than the fundamental scale $M_D$, i.e.
$f\ll M_D$, the KK modes decouple from the SM particles. Therefore, 
for flexible enough branes, the only relevant degrees of freedom at
low energies in the BWS are the SM particles and the branons.
Similarly to other GB's, branons are expected to be nearly
massless and weakly interacting at low energies. Nevertheless, 
in general, translational invariance in the extra dimensions
is not necessarily an exact symmetry and 
some
branon mass $M$ is expected from such explicit symmetry breaking
as shown in \cite{DoMa,ACDM}. This is similar to what happens
to pions which are the GB
corresponding to the SSB of the chiral symmetry of low-energy
strong interactions. As gravitons do, branons couple to
$T_{SM}^{\mu\nu}$, however in this case, the lowest order
effective Lagrangian is \cite{ACDM}:
\begin{eqnarray}
{\cal L}_{Br}&=&
\frac{1}{2}g^{\mu\nu}\partial_{\mu}\pi^\alpha
\partial_{\nu}\pi^\alpha-\frac{1}{2}M^2\pi^\alpha\pi^\alpha\nonumber  \\
&+& \frac{1}{8f^4}(4\partial_{\mu}\pi^\alpha
\partial_{\nu}\pi^\alpha-M^2\pi^\alpha\pi^\alpha g_{\mu\nu})
T^{\mu\nu}_{SM}\label{lag}
\end{eqnarray}
We see that branons always interact by pairs, 
they are stable and difficult to detect,  since
their interactions are suppressed by the  tension scale $f$, 
and they are expected to be massive. Thus
we arrive to the conclusion that the massive oscillations of the
brane are natural candidates to dark matter in the BWS where $f\ll
M_D$. The dark matter problem has
been considered in this scenario from a different point of 
view in \cite{Dvali}
and also in models with universal extra dimensions \cite{CFM}.

In order to calculate the thermal relic branon abundance, we will
use the standard techniques given in \cite{Kolb} in two 
limiting cases,
either  relativistic (hot) or non-relativistic (cold) 
branons at decoupling.
The evolution of the number density $n_\alpha$ of branons
 $\pi^\alpha$,
 $\alpha=1,\dots , N$  interacting with SM particles in an 
expanding universe is given
by the Boltzmann equation:
\begin{eqnarray}
\frac{dn_\alpha}{dt}=-3Hn_\alpha-\langle \sigma_A v\rangle
(n_\alpha^2 -(n_\alpha^{eq})^2)\label{Boltzmann}
\end{eqnarray}
where $\sigma_A=\sum_X \sigma(\pi^\alpha\pi^\alpha\rightarrow X)$
is the total annihilation cross section of branons into SM
particles $X$ summed over all final states. The $-3Hn_\alpha$
term, with $H$ the Hubble parameter, takes into account the
dilution of the number density due to the universe expansion.
These are the only terms which could change their number density
to the leading order.
In fact, branons
 do not decay into other particles and since they
interact always by pairs the conversions like $\pi^\alpha X
\rightarrow \pi^\alpha Y$ do not change their number. Notice that
we are considering the low-energy effective 
Lagrangian in (\ref{lag})
and assuming for simplicity that all the  branons are degenerate.
Accordingly, each
branon species evolves independently, and in the following
we will drop the $\alpha$ index. The total branon density will be
just $N$ times that of a single branon. The $\langle \sigma_A
v\rangle$ term denotes the
 thermal average of the total annihilation cross section
times the relative velocity. From (\ref{lag}),
it includes, to leading order, annihilations into all 
the SM particle-antiparticle pairs.
If the universe temperature is above the
QCD phase transition ($T>T_c$),
we consider  annihilations into quark-antiquark and gluons pairs. 
If $T<T_c$  we 
include annihilations
into light hadrons. For the sake of definiteness we have taken a
critical temperature $T_c\simeq 170$ MeV and a Higgs mass $m_H\simeq
125$ GeV, although the final results are not very sensitive to the
concrete value of these parameters.

Defining the new variable $x=M/T$, one
finds from (\ref{Boltzmann}) that below the freeze-out 
temperature $x_f$ for which the
annihilation rate  $\Gamma_A=n_{eq}\langle\sigma_A v
\rangle$  equals the expansion rate  $H$, 
branons are decoupled from the thermal bath and their abundance
remains frozen relative to the entropy
density.
We will denote $g_{eff}(T)$ and $h_{eff}(T)$  the effective 
relativistic
degrees of freedom for energy and entropy densities 
respectively at temperature $T$.
In order to calculate the decoupling temperature 
for hot relics, 
it is a good approximation to use the condition
$\Gamma_A(x_f) \simeq H(x_f)$. From the Friedmann equation
in a radiation dominated universe we have: 
$H(x)\simeq 1.67 \;g_{eff}^{1/2} \;M^2/(x^2\, M_P)$
On the other hand,  expanding $\Gamma_A(x)$ for $x \ll 3$ 
and  neglecting  $M$, we find: 
$\Gamma_A^\gamma(T)\simeq 16\pi^9T^9/(297675\zeta(3)f^8)$ for
photons and $\Gamma_A^\nu(T)=\Gamma_A^\gamma(T)/4$ for neutrinos.
For massive particles we cannot  give  closed expressions.
Once we know $x_f$, the corresponding fraction of
energy density today in the form of relics
is given by:
\begin{eqnarray}
\Omega_{Br} h^2 \simeq \frac{7.83 \cdot 10^{-2}}{h_{eff}(x_f)}\,
\frac{M}{\mbox{eV}}\label{eV}
\end{eqnarray}
For  cold
relics the calculation of   the decoupling temperature is  more
complicated. The well-known result is:
\begin{eqnarray}
x_f=\ln\left(\frac{0.038\, c\,(c+2) M_P M \langle\sigma_A v
\rangle}{g_{eff}^{1/2}\,x_f^{1/2}}\right)
\label{xf}
\end{eqnarray}
where $c\simeq 0.5$ is obtained from the numerical resolution
of the Boltzmann equation.
The above equation can be solved iteratively. The  
matter density
can be written as:
\begin{eqnarray}
\Omega_{Br} h^2 \simeq 8.766 \cdot 10^{-11} \mbox{GeV}^{-2}
\frac{x_f}{g_{eff}^{1/2}}\left(\sum_{n=0}^\infty
\frac{c_n}{n+1}x_f^{-n}\right)^{-1}
\label{coldomega}
\end{eqnarray}
where we have expanded $\langle \sigma_A v\rangle $ in powers of
$x^{-1}$ as $\langle \sigma_A v\rangle =\sum_{n=0}^\infty c_n
x^{-n}$. In the case of photons, the first
non-vanishing coefficient is $c_2^\gamma=68 M^6/(15f^8 \pi^2)$ and for
massless neutrinos $c_2^\nu=c_2^\gamma/4$ ($d-$wave annihilation)
whereas for non-conformal matter we also have $s-$ and $p-$wave annihilation,
($c_0,c_1\neq 0$). The corresponding
expressions are more complicated and will be given elsewhere. We
have performed all the expansions up to $\Od(x^{-2})$.
Coannihilation effects are absent in this case  since there are no
slightly heavier particles which eventually could decay into the
lightest branon. Also, in order to avoid the problems of the
Taylor expansion near SM thresholds, we have taken branon masses
sufficiently separated from SM particles masses where the usual
treatment is adequate \cite{Kolb}. Such treatment 
is known to introduce errors of the order of 10$\%$ in the
relic abundances.

\begin{figure}[h]
\resizebox{8.5cm}{!}{\includegraphics{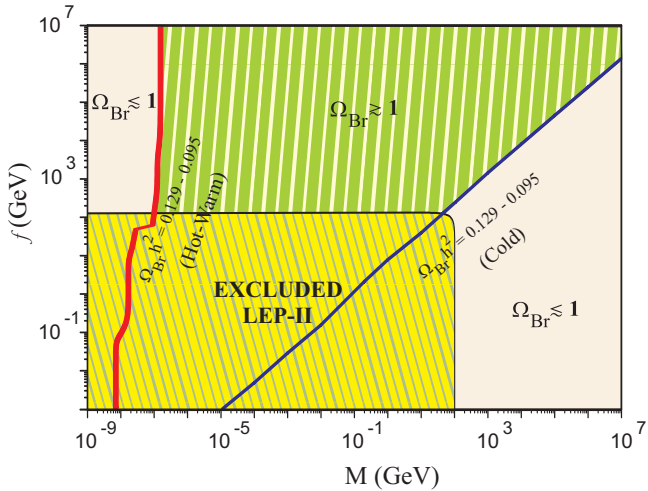}}
\caption{Relic abundance for a model
with one branon $N=1$. The line on the left
is the $\Omega_{Br}h^2=0.129 - 0.095$ curve for hot-warm relics. 
The 
line on the right corresponds to cold relics. 
The lower striped area is the estimated 
excluded region by single-photon
processes at LEP-II \cite{ACDM} 
and the upper area  is also excluded by branon overproduction.}
\end{figure}

Branons could be responsible for the observed cosmological dark
matter density provided $\Omega_{Br}h^2 = 0.129 - 0.095$ 
at the 95$\%$ C.L. which
corresponds to $\Omega_{Br}=0.23 \pm 0.08$ and $h=0.79-0.65$ 
\cite{WMAP}.
 In FIG. 1, we have plotted these curves for  hot and cold
branons in the $f-M$ plane for one single branon. For $N$ types of
branons
the corresponding abundances are simply $N$ times larger.
In fact,
the contribution of branons to $g_{eff}$
in (\ref{coldomega})
is negligible in the case of cold relics. For hot relics
 such
contribution in (\ref{eV}) has been taken into account,
although it is very small in the interesting regions.
Concerning the freeze-out temperature,
the results on the cold dark matter curve range from $x_f\simeq 9$
for $M=10^{-5}$ GeV to $x_f\simeq 31$  for $M=10^6$ GeV.
In the hot relics case, a very good approximation  is given by
$\log (T_f/\mbox{GeV})\simeq (8/7)\log (f/\mbox{GeV})-3.2$.

Pure hot dark matter
models are disfavored at present because relativistic 
matter free streams from overdense into
underdense regions preventing structures from growing below the so
called free-streaming scale given by \cite{Colin}:
$\lambda_{FS}\simeq 0.2\, (\Omega_{Br}h^2)^{1/3}
(\mbox{keV}/M)^{4/3}\,\mbox{Mpc}$.
In the hot branons curve in FIG. 1, the masses in the
allowed region are in the range
$M=85-177$ eV, which corresponds to 
$\lambda_{FS}\simeq 2.4-1.0$ Mpc. Such scales are much
smaller than those in neutrino dark matter models and, in addition,
since these branons decouple much earlier than neutrinos
do,
their corresponding temperatures are also lower, i.e. they 
could be considered rather as
warm dark matter (WDM) candidates \cite{Colin} from the point 
of view of
structure formation. However, a WDM dominated universe
seems to be also disfavored by the 
recent observations of WMAP \cite{WMAP}. 
On the other hand, the presence of relativistic
branons during BBN could
change the expansion rate of the universe, spoiling the
predictions of the light elements abundances \cite{Kolb}.  
However, the BBN
limits on the number of branons contributing 
to $g_{eff}$ are
not very constraining. In fact, imposing the conservative
bound  $g_{eff}(T_{BBN}\sim 1 \,\mbox{MeV})\lsim 12.5$ \cite{Kolb}, 
we find $N \lsim 18$ for branons which decoupled
before the QCD phase transition, corresponding to $f\gsim 60$ GeV. 

Let us clarify the main 
assumptions  used so far in the work.
We are considering models with $f\ll M_D$, and also
assuming that  the evolution of
the universe is standard up to a temperature around $f$.
Indeed, this is the case of realistic brane cosmology models
\cite{Langlois}. Therefore, we have taken 
the conservative bound $f>1$ MeV, 
so that BBN is not affected. 
Moreover the 
effective Lagrangian (\ref{lag}) is only valid at low energies 
relative
to  $f$. 
We have checked that our calculations are consistent with these
assumptions 
since the decoupling temperatures   
are always smaller than $f$ in the allowed regions in FIG. 1.

Brane fluctuations could be, not only  candidates for the cosmological
dark matter, but also they could make up the galactic halo and explain
the local dynamics. In such case, they could be detected in
direct search experiments from the energy transfer in elastic collisions
with nuclei of a suitable target. The appropriate quantity to be compared
with the  experimental results  is not the 
elastic branon-nucleus cross section $\sigma$,
 but  the differential
cross section per nucleon at zero-momentum transfer $\sigma_n$,
which is defined by \cite{Kamion}:
%\begin{eqnarray}
%\frac{d\sigma}{d|q|^2}=\frac{\sigma_n A^2\,F^2(|q|)}{4v^2\mu^2},\,\,
%\mbox{with}\,\,\sigma_n=\frac{9M^2m^2\mu^2}{64\pi f^8}
%\end{eqnarray}
\begin{eqnarray}
\frac{d\sigma}{d|q|^2}=\frac{\sigma_n A^2\,F^2(|q|)}{4v^2\mu^2}
\end{eqnarray}
where $\mu=Mm/(M+m)$,  $F(|q|)$ is a nuclear form factor 
with the normalization $F(0)=1$, $m\simeq 939$ MeV
is the nucleon mass, $v$ is the relative velocity and 
$A$ is the mass number
of the nucleus. In the limit
in which the momentum transfer goes to zero, we can
consider the nucleons as
pointlike particles.
In this case, it  is possible to calculate the 
 branon-nucleon cross section
$\sigma_n$
from (\ref{lag}) just considering the nucleon as a Dirac 
fermion of mass $m$:
\begin{eqnarray}
\sigma_n=\frac{9M^2m^2\mu^2}{64\pi f^8}
\end{eqnarray}
In fact, this quantity does not depend on the type of particle
which couples to the branon, but only on its mass. This can 
be seen from
(\ref{lag}) since in this limit, branons only couple to the 
$T^{00}$
component.
\begin{figure}[h]
\resizebox{8.5cm}{!}{\includegraphics{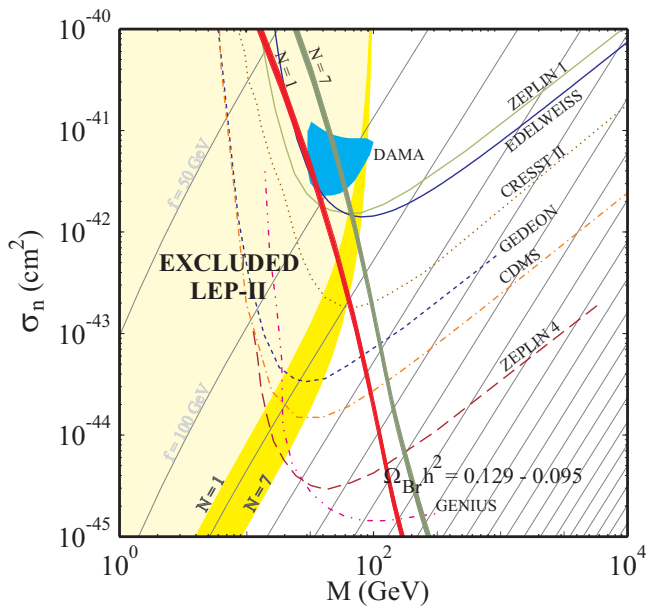}}
\caption{ Elastic branon-nucleon cross section 
$\sigma_n$ in terms of the branon
mass. The two thick lines correspond to the 
 $\Omega_{Br}h^2 = 0.129 - 0.095$ curve for cold branons 
in Fig.1 with $N=1$ (left) and $N=7$ (right). The shaded 
areas on the left are the previous LEP-II
exclusion regions \cite{ACDM}, 
also for $N=1,7$. The solid lines correspond to the current limits
on spin-independent cross section from direct detection 
experiments: ZEPLIN1 \cite{Zeplin1},
DAMA \cite{Dama}
and EDELWEISS \cite{Edelweiss}. The discontinuous lines are the projected 
limits for:
CRESST \cite{Cresst}, GEDEON \cite{Gedeon}, CDMS \cite{CDMS}, 
ZEPLIN4 \cite{Zeplin4} and
GENIUS \cite{Genius}, (limits obtained from \cite{dmtools}).}
\end{figure}

The results of our analysis  are shown in FIG. 2. Lines 
of constant $f$
 with 50 GeV separation are shown for reference.  
The area
on the left of the
$\Omega_{Br} h^2=$$0.129 - 0.095$ curves is excluded by branon overproduction, but
the right portion is compatible with  observations
and will be explored in  future experiments. Such region corresponds
to $f \gsim 120$ GeV and $M \gsim 40$ GeV.

Another interesting possibility is to detect  branons indirectly. 
Their annihilations
in the galactic halo can give rise to pairs of photons or 
$e^+e^-$ which could
be detected by 
$\gamma -$rays telescopes such as MAGIC or GLAST
or antimatter detectors (AMS). Annihilation 
of branons trapped in the center of the sun or the
earth can give rise to high-energy neutrinos which could be
detectable by high-energy neutrino telescopes such 
as AMANDA, IceCube or  ANTARES
(see for example \cite{CFM}). 
Because annihilations of non-relativistic branons 
into conformal matter are $d-$wave suppressed, the most relevant 
contribution will
come from the secondary leptonic decays of ultra-relativistic 
$Z$ or $W$ bosons (in the case $M\gg M_Z$).
These processes will be characterized by the presence of peaks around
one half of the branon mass in the leptonic or neutrino spectra.
In the case of photons, softer peaks will be present at lower energies
and therefore their detection will be more difficult.
The hadronic decays will give rise to  relatively smeared spectra
at lower energies. Detailed results will be presented elsewhere.

Throughout the paper we have assumed that branons 
were in thermal equilibrium 
with radiation at some point in the history of the universe.
If this is not the case, branons could still be produced 
non-thermally, very much  
in the same way as  axionic dark matter \cite{Wilczek}. 
In fact for very light branons, the energy density produced by this mechanism
could be cosmologically important.

In conclusion, we have proposed branons as  natural dark matter candidates in
the BWS with low tension. Our results show that in a certain range
of the parameters $f$ and $M$, their relic abundances could 
explain the
missing mass problem, and that such parameters region will
be explored in future direct detection experiments.

 {\bf Acknowledgements:} This work
 has been partially supported by the DGICYT (Spain) under the
 project numbers FPA 2000-0956 and BFM2000-1326.
%\bibliography{DM12}% Produces the bibliography via BibTeX.

\end{document}